\documentclass{webofc}
\usepackage[varg]{txfonts}   % Web of Conferences font
\begin{document}
\title{The \textsc{Majorana Demonstrator} Status and Preliminary Results}
%
% subtitle is optional

\author{\firstname{C.-H.} \lastname{Yu}\inst{1} \and
        \firstname{S.I.} \lastname{Alvis}\inst{2}\and
         \firstname{I.J.} \lastname{Arnquist}\inst{3}\and
         \firstname{F.T.} \lastname{Avignone III}\inst{1,4}\and
         \firstname{A.S.} \lastname{Barabash}\inst{5}\and
          \firstname{C.J.} \lastname{Barton}\inst{6}\and
           \firstname{F.E.} \lastname{Bertrand}\inst{1}\and
           \firstname{T.} \lastname{Bode}\inst{7}\and  
            \firstname{V.} \lastname{Brudanin}\inst{8}\and   
            \firstname{M.} \lastname{Busch}\inst{9,10}\and   
            \firstname{M.} \lastname{Buuck}\inst{2}\and
            \firstname{T.S.} \lastname{Caldwell}\inst{9,11}\and      
            \firstname{Y.-D.} \lastname{Chan}\inst{12}\and
            \firstname{C.D.} \lastname{Christofferson}\inst{13}\and
             \firstname{P.-H.} \lastname{Chu}\inst{14}\and
             \firstname{C.} \lastname{Cuesta}\inst{2}\fnsep\thanks{Present Address: Centro de Investigaciones Energ{\'e}ticas, Medioambientales y Tecnol{\'o}gicas, CIEMAT, 28040, Madrid, Spain}
             \firstname{J.A.} \lastname{Detwiler}\inst{2}\and
            \firstname{C.} \lastname{Dunagan}\inst{13}\and
            \firstname{Yu} \lastname{Efremenko}\inst{1,15}\and
            \firstname{H.} \lastname{Ejiri}\inst{16}\and 
            \firstname{S.R.} \lastname{Elliott}\inst{14}\and
             \firstname{T.} \lastname{Gilliss}\inst{9,11}\and
            \firstname{G.K.} \lastname{Giovanetti}\inst{17}\and
            \firstname{M.} \lastname{Green}\inst{1,9,18}\and
            \firstname{J.} \lastname{Gruszko}\inst{19}\and
            \firstname{I.S.} \lastname{Guinn}\inst{2}\and
            \firstname{V.E.} \lastname{Guiseppe}\inst{4}\and
            \firstname{C.R.} \lastname{Haufe}\inst{9,11}\and
            \firstname{L.} \lastname{Hehn}\inst{12}\and
            \firstname{R.} \lastname{Henning}\inst{9,11}\and
            \firstname{E.W.} \lastname{Hoppe}\inst{3}\and
            \firstname{M.A.} \lastname{Howe}\inst{9,11}\and
            \firstname{K.J.} \lastname{Keeter}\inst{20}\and
            \firstname{M.F.} \lastname{Kidd}\inst{21}\and
            \firstname{S.I.} \lastname{Konovalov}\inst{5}\and
            \firstname{R.T.} \lastname{Kouzes}\inst{3}\and
            \firstname{A.M.} \lastname{Lopez}\inst{15}\and
            \firstname{R.D.} \lastname{Martin}\inst{22}\and
            \firstname{R.} \lastname{Massarczyk}\inst{14}\and
            \firstname{S.J.} \lastname{Meijer}\inst{9,11}\and
             \firstname{S.} \lastname{Mertens}\inst{7,23}\and 
              \firstname{J.} \lastname{Myslik}\inst{12}\and
             \firstname{G.} \lastname{Othman}\inst{9,11}\and
              \firstname{W.} \lastname{Pettus}\inst{2}\and
               \firstname{A.W.P.} \lastname{Poon}\inst{12}\and
              \firstname{D.C.} \lastname{Radford}\inst{1}\and
               \firstname{J.} \lastname{Rager}\inst{9,11}\and
               \firstname{A.L.} \lastname{Reine}\inst{9,11}\and
               \firstname{K.} \lastname{Rielage}\inst{14}\and
                \firstname{N.W.} \lastname{Ruof}\inst{2}\and     
                 \firstname{B.} \lastname{Shanks}\inst{1}\and  
                 \firstname{M.} \lastname{Shirchenko}\inst{8}\and
                  \firstname{A.M.} \lastname{Suriano}\inst{13}\and
                  \firstname{D.} \lastname{Tedeschi}\inst{4}\and
                  \firstname{R.L.} \lastname{Varner}\inst{1}\and
                   \firstname{S.} \lastname{Vasilyev}\inst{8}\and
                   \firstname{K.} \lastname{Vetter}\inst{12}\fnsep\thanks{Alternative address:  Department of Nuclear Engineering, University of California, Berkeley, CA, USA}\and  
                    \firstname{K.} \lastname{Vorren}\inst{9,11}\and
                     \firstname{B.R.} \lastname{White}\inst{14}\and
                     \firstname{J.F.} \lastname{Wilkerson}\inst{1,9,11}\and
                     \firstname{C.} \lastname{Wiseman}\inst{4}\and
                       \firstname{W.} \lastname{Xu}\inst{6}\and
                       \firstname{E.} \lastname{Yakushev}\inst{8}\and
                       \firstname{V.} \lastname{Yumatov}\inst{5}\and
                      \firstname{I.} \lastname{Zhitnikov}\inst{8}\and
                     \firstname{B.Z.} \lastname{Zhu}\inst{14}
\center {(\textsc{Majorana} Collaboration)}
}

\institute{Oak Ridge National Laboratory, Oak Ridge, TN, USA
\and
            Center for Experimental Nuclear Physics and Astrophysics, and Department of Physics, University of Washington, WA, USA
\and
	    Pacific Northwest National Laboratory, Richland, WA, USA
\and
	   Department of Physics and Astronomy, University of South Carolina, Columbia, SC, USA   	    
\and
           National Research Center, ``Kurchatov Institute'' Institute for Theoretical and Experimental Physics, Moscow, Russia
\and
	Department of Physics, University of South Dakota, Vermillion, SC, USA
\and
	Max-Planc-Institute f{\"u}r Physik, M{\"u}nchen, Germany
\and
	Joint Institute for Nuclear Research, Dubna, Russia
\and
	Triangle Universities Nuclear Laboratory, Durham, NC, USA
\and
	Department of Physics, Duke University, Durham, NC, USA
\and
	Department of Physics and Astronomy, University of North Carolina, Chapel Hill, NC, USA
\and
	Nuclear Science Division, Lawrence Berkeley National Laboratory, Berkeley, CA, USA
\and
	South Dakota School of Mines and Technology, Rapid City, SD, USA
\and
	Los Alamos National Laboratory, Los Alamos, NM, USA
\and
	Department of Physics and Astronomy, University of Tennessee, Knoxville, TN, USA
\and
	Research Center for Nuclear Physics, Osaka University, Ibaraki, Osaka, Japan
\and
	Department of Physics, Princeton University, Princeton, NJ, USA
\and
	Department of Physics, North Carolina State University, Raleigh, NC, USA
\and
	Department of Physics, Massachusetts Institute of Technology, Cambridge, MA, USA
\and
	Department of Physics, Black Hills State University, Spearfish, SD, USA
\and
	Physics Department, Tennessee Tech University, Cookeville, TN, USA
\and
	Department of Physics, Engineering and Astronomy, Queen's University, Kingston, ON, Canada
\and
	Physik Department, Technische Universit{\"a}t, M{\"u}nchen, Germany
}

\abstract{
The \textsc{Majorana} Collaboration is using an array of high-purity Ge detectors to search for neutrinoless double-beta decay in $^{76}$Ge. 
Searches for neutrinoless double-beta decay are understood to be the only viable experimental method for testing the Majorana nature of the neutrino. 
Observation of this decay would imply violation of lepton number, that neutrinos are Majorana in nature, and provide information on the neutrino mass. 
The \textsc{Majorana Demonstrator} comprises 44.1 kg of p-type point-contact  Ge detectors (29.7 kg enriched in $^{76}$Ge) surrounded by a low-background shield system.  The experiment achieved a high efficiency of converting raw Ge material to detectors and an unprecedented detector energy resolution of 2.5 keV FWHM at $Q_{\beta\beta}$. The \textsc{Majorana} collaboration began taking physics data in 2016.
This paper summarizes key construction aspects of the \textsc{Demonstrator} and shows preliminary results from initial data.  
  
  }
\maketitle

\section{Introduction}

Neutrinoless double-beta (0$\nu\beta\beta$) decay experiments play an important  role in understanding the nature of neutrinos and their connection with the evolution of the universe. The observation of 0$\nu\beta\beta$ decay would indicate that lepton number is not conserved -- a key ingredient of thermal leptogenesis --  and that neutrinos are Majorana particles \cite{Zralek, Schechter}.
Measurements of the 0$\nu\beta\beta$  decay rates would also yield information on the absolute neutrino mass.  For more comprehensive reviews of 0$\nu\beta\beta$ decays, see Refs. \cite{Avignone, Barabash, Rodejohann, Elliott1, Vergados, Cremonesi, Schwingenheuer, Elliott2, Henning}.

The \textsc {Majorana Demonstrator}  is an experiment using  enriched- and natural Ge as the detector and source to search for neutrinoless double-beta decay from $^{76}$Ge. The primary goal of the \textsc{Demonstrator} is to build a scalable experiment achieving low enough backgrounds to justify the feasibility of a larger-scale experiment.  The following sections  give a short summary of the design, operation, data analysis and initial results from the \textsc {Majorana Demonstrator}.

\section{The \textsc{Majorana Demonstrator}}

By choosing the isotope $^{76}$Ge for the 0$\nu\beta\beta$ search, the \textsc{Majorana Demonstrator} is able to take advantage of the well-established Ge detector technology to achieve excellent energy resolution and event reconstruction (using pulse shape analysis) with intrinsically low backgrounds.  For the \textsc{Demonstrator} , the average  FWHM  at the $Q_{\beta\beta}$ energy is  $2.5\pm 0.1$ keV, the best energy resolution achieved to date for a 0$\nu\beta\beta$ decay experiment.

A total of 44.1 kg of Ge was used for the \textsc {Majorana Demonstrator}, 
29.7 kg of which is enriched to 88\% of isotope 76.    The p-type point-contact  (PPC)  detector design \cite{Luke, Barbeau, Aguayo} provides effective single-site/multi-site event discrimination, low noise, and low-energy thresholds that make it possible to search for new physics at low energies.   To fabricate the Ge detectors, the enriched raw $^{76}$GeO$_2$ material was reduced, zone refined, grown into single crystals, and then made into 35 detectors.  Overall, the \textsc{Majorana Demonstrator} reached an efficiency of 69.8\% in converting the initial $^{76}$GeO$_2$ material into Ge detectors, the highest efficiency achieved to date \cite{Abgrall1}. The 35 detectors were assembled into 14 strings (3--5 detectors in each string) and installed in two cryostats with 7 strings in each.  Care was taken to minimize the cosmogenic activation by limiting the total surface exposure time of these detectors.  On average, the effective sea-level surface exposure time of the enriched detectors was approximately 23 days.

\begin{figure}[h]
\centering
\includegraphics[width=8cm,clip]{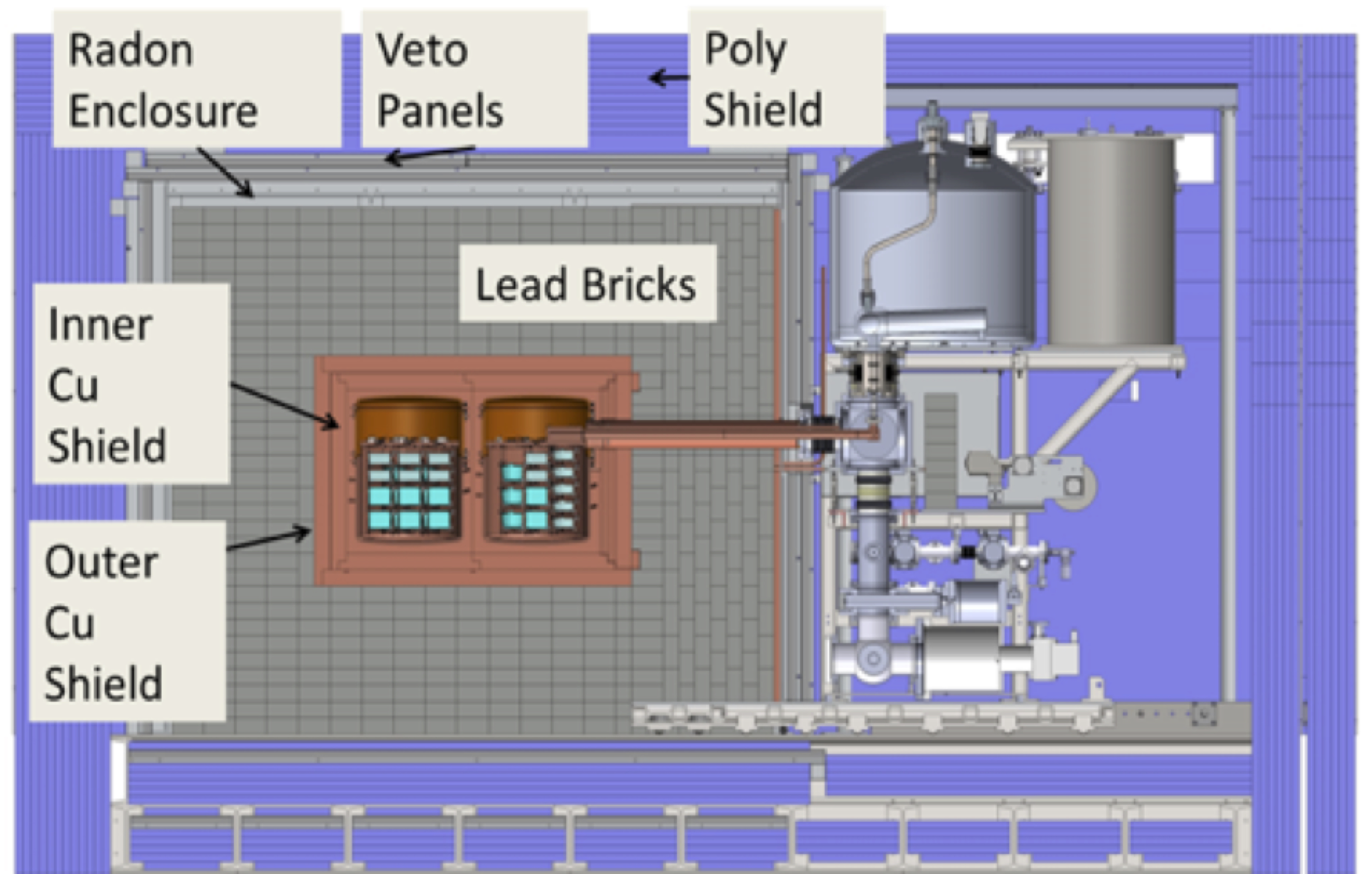}
\caption{An illustration of the \textsc{Majorana Demonstrator}  shield system in cross section. Detectors are denoted by blue disks stacked in strings installed in two electroformed copper cryostats.  The cryostats are surrounded by electroformed inner- and clean-commercial outer copper shields. Lead bricks, a radon enclosure, muon veto panels, and polyethylene shields complete the shield system.}
\label{fig-1}      
\end{figure}

To reject external background, the detector cryostats are surrounded by multi-layer shields and installed in the clean lab at the Sanford Underground Research Facility (SURF) in Lead, South Dakota, USA at 4850 feet below the surface.   Figure~\ref{fig-1} shows a cross sectional illustration of the \textsc{Demonstrator}.  Starting from the center cavity, two cryostat modules constructed from electroformed copper house the detector strings.  The cryostat cavity is surrounded by an electroformed copper inner shield  and an outer copper shield constructed from clean commercial copper.  Outside the copper shields is a thick layer of high-purity lead. A muon veto system consisting of 32 aluminum-clad plastic scintillator panels surrounds the lead and copper shields.  The muon veto was used to measure the muon flux in the underground clean lab.  The measured  flux of $5\times 10^{-9} \mu/s/cm^2$ is in good agreement with simulations\cite{Abgrall2}. To reduce neutron background, a polyethylene outer shield completes the shield system.

To reduce intrinsic background from sources near the detectors, careful designs enabled the least amount of material for detector assembly, including front-end circuit boards, connectors, cables, and string parts. 
Copper was chosen as the material for the \textsc{Demonstrator} cryostats and inner shielding layers due to its lack of naturally-occurring radioactive isotopes, as well as its suitability for machining.  Electroforming further reduces radioactivity from contaminants.   To minimize surface exposure of material, an electroforming lab, a machine shop, and part-cleaning facilities were established underground at SURF.  As a result, most copper components installed in the \textsc{Demonstrator}  have no cosmogenic activation from surface-level cosmic rays.

\section{Data Accumulation and Analysis}
\label{sec-data}

The assembly of the first module (cryostat) of the \textsc{Demonstrator} was completed in 2015.  This module consists of 16.8 kg of enriched Ge detectors and 5.6 kg of natural Ge detectors.  Data taking of Module 1 alone began in June 2015.  The assembly of Module 2 with 12.9 kg of enriched Ge and 8.8 kg natural Ge  was completed in August 2016.  The full operation of the \textsc{Demonstrator} started with both modules taking data in August 2016.

Since the start of operation up to this presentation,  several data sets (DS) have been collected, corresponding to various experimental setup and running conditions.  Before the installation of the inner copper shield, a serious of commissioning runs were carried out, and these were collected in DS0.  After the  inner copper shield was installed, data were collected in DS1.  Subsequently, a series of tests were performed on signal multisampling of the digitized waveforms  in order to extend  the signal capture after an event, therefore making it more effective to discriminate against $\alpha$ surface-background events.  These data were collected in DS2. After the implementation of  signal multisampling,  Module 1 and Module 2 data were respectively collected into DS3 and DS4 with separate data acquisition systems.  Results presented at this conference are those from DS3 and DS4, which were the first two datasets with detector-, shield-, and data-acquisition configurations that are close to the final configuration.

  The collected data were first scanned to remove non-physical waveforms.  Then, events that are associated with multi-site energy deposits were removed by coincidence-event rejection and pulse-shape analysis\cite{Cuesta}.  Double-beta decay events are considered single-site events because the range of the electron in Ge is small compared to that of the Compton-scattered $\gamma$ events, which are backgrounds.  Finally, events associated with external $\alpha$ energy depositions on the passivated surface of detectors were rejected by analysis cuts. Extensive tests indicate that these $\alpha$ activities were most likely due to $^{210}$Pb-supported $^{210}$Po as daughters of Rn contamination on the Teflon components of the detector mount.  For most of these $\alpha$ events, charge drifting takes place along the detector surface, whereas for most non-$\alpha$ events,  charge drifting takes place inside the bulk of the detector.  The difference of surface- and bulk-drifting times enables the rejection of $\alpha$ events by pulse shape analysis.

\section{Preliminary Results}

\begin{figure}[h]
\centering
\includegraphics[width=9 cm,clip]{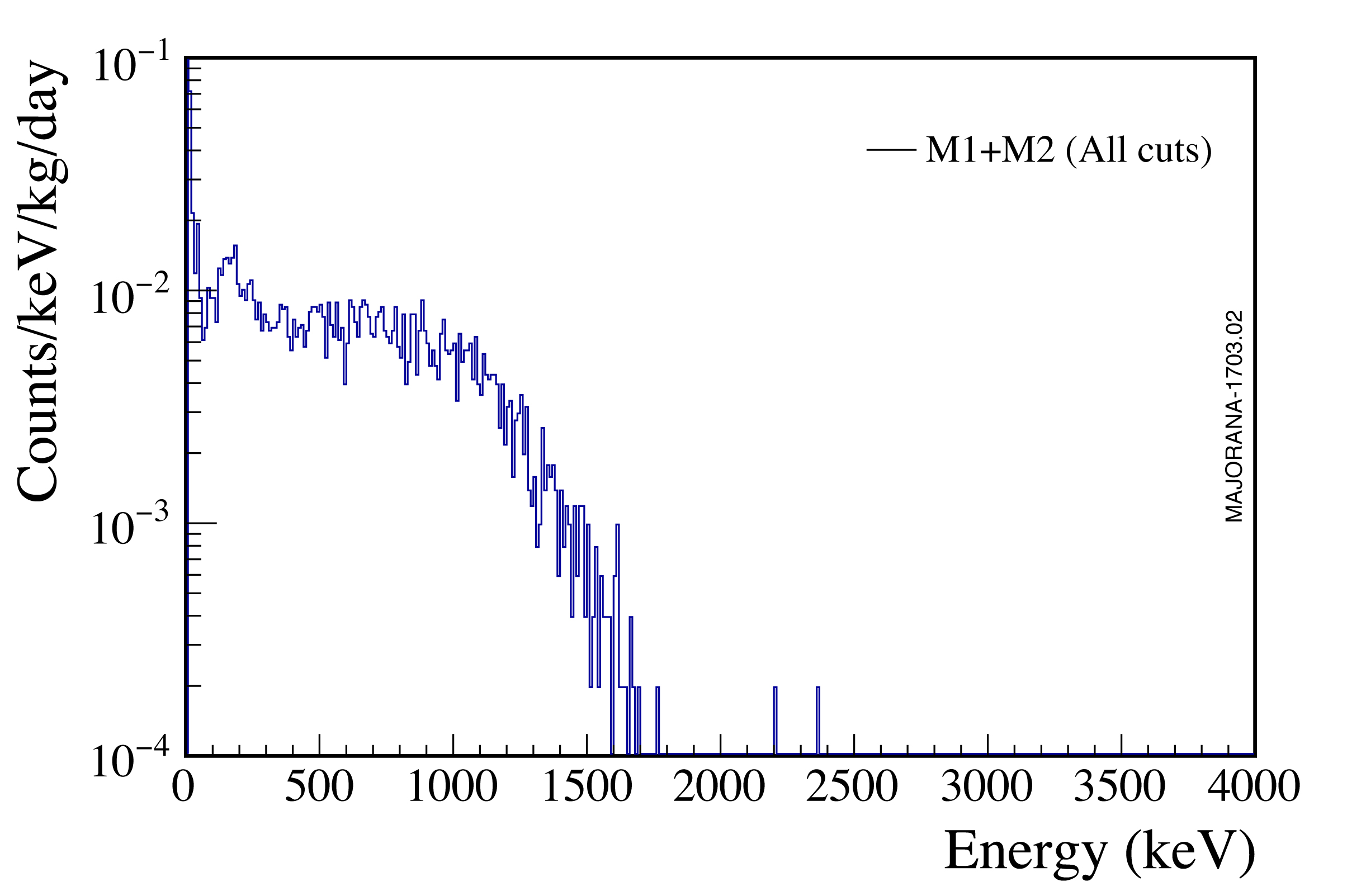}
\caption{Summed background spectrum from DS3 and DS4 after all cuts (enriched Ge detectors only). One count remains in the 400-keV region of interest centered at the 2039-keV 0$\nu\beta\beta$ Q value.   }
\label{fig-2}       % Give a unique label
\end{figure}

By the fall of 2017, the \textsc{Demonstrator} had collected more than 20 kg yr of data.  Of these data,  the first 1.39 kg yr of enriches-Ge data collected with both modules is presented  in this paper. A summed spectrum of enriched-Ge detectors from DS3 and DS4  is shown in Figure~\ref{fig-2}.  This spectrum is the result of applying all analysis cuts, including pulse-shape discrimination (PSD) and $\alpha$-contamination cuts.  After all cuts, 1 count remains in the 400-keV region of interest centered at the 2039-keV 0$\nu\beta\beta$ Q value.  This translates into a background rate of $5.0^{+8.8}_{-3.2}$ c/(ROI t yr) at 90\% confidence level,  where the ROI (region of interest) is a 2.8-keV window centered at the 0$\nu\beta\beta$ Q value, or a background index of $1.8 ^{+3.1}_{-1.1} \times 10^{-3}$ c/(keV kg yr). 

It should be noted that at the time of this proceeding, the \textsc{Majorana Demonstrator} has more than 7 times data collected and analyzed than what is shown in this presentation.    For a current description of results from the \textsc{Majorana Demonstrator}, see Ref.~\cite{Aalseth}.

Although data analyzed so far corresponds to limited exposure, the \textsc{Majorana Demonstrator} is approaching half-life limits comparable to the best effort to date, due to the very low background and excellent energy resolution of the experiment. Both the \textsc{Majorana Demonstrator} and GERDA are currently operating in the nearly-background free regime.  The latest results from both experiments point to a $^{76}$Ge half-life limit approaching $10^{26}$ yr\cite{Agostini,Aalseth},  and indicate that a future large-scale $^{76}$Ge experiment, such as LEGEND \cite{Abgrall3}, is indeed warranted.

\section*{Acknowledgments}
This material is based upon work supported by the U.S. Department of Energy, Office of Science, Office of Nuclear Physics, the Particle Astrophysics and Nuclear Physics Programs of the National Science Foundation, the Russian Foundation for Basic Research, the Natural Sciences and Engineering Research Council of Canada, the Canada Foundation for Innovation John R.~Evans Leaders Fund, the National Energy Research Scientific Computing Center, the Oak Ridge Leadership Computing Facility, and the Sanford Underground Research Facility.  This research used resources of the Oak Ridge Leadership Computing Facility, which is a DOE Office of Science User Facility supported under Contract DE-AC05-00OR22725.

\end{document}